\documentclass[twocolumn,twocolappendix]{aastex63}
\usepackage{amsmath}
\usepackage{amssymb}
\usepackage{bm}
\usepackage{graphicx}
\usepackage{color}
\usepackage{float}
\usepackage{tabularx}
\usepackage{ragged2e}
\usepackage{graphicx}
\usepackage{subfigure}

\definecolor{darkgreen}{rgb}{0,0.35,0}
\definecolor{blue}{rgb}{0,0,1}
\usepackage{CJKutf8}            % Chinese name

\newcommand{\be}{\begin{eqnarray}}
\newcommand{\ee}{\end{eqnarray}}

\shorttitle{Migration of Gas Accreting Planets}
\shortauthors{Ida et al.}

\begin{document}

\title{Outward Migration of a Gas Accreting Planet: A Semi-Analytical Formula}

\author[0000-0002-9676-3891]{Shigeru Ida \begin{CJK*}{UTF8}{gbsn}(井田茂)\end{CJK*}}
\affiliation{Earth-Life Science Institute, Institute of Science Tokyo, Tokyo 152-8550, Japan}
\affiliation{Department of Astronomy, School of Science, Westlake University, Hangzhou, Zhejiang 310030, China}
\author[0000-0002-7329-9344]{Ya-Ping Li \begin{CJK*}{UTF8}{gbsn}(李亚平)\end{CJK*}}
\affiliation{Shanghai Astronomical Observatory, Chinese Academy of Sciences, Shanghai 200030, China}
\author[0009-0002-7003-8883]{Jun-Peng Pan \begin{CJK*}{UTF8}{gbsn}(潘俊鹏)\end{CJK*}}
\affiliation{Shanghai Astronomical Observatory, Chinese Academy of Sciences, Shanghai 200030, China}
\affiliation{University of Chinese Academy of Sciences, Beijing 100049, China}
\author[0000-0003-3792-2888]{Yi-Xian Chen \begin{CJK*}{UTF8}{gbsn}(陈逸贤)\end{CJK*}}
\affiliation{Department of Astrophysics, Princeton University, Princeton, NJ 08544, USA}
\author[0000-0001-5466-4628]{Douglas N. C., Lin \begin{CJK*}{UTF8}{gbsn}(林潮)\end{CJK*}}
\affiliation{Department of Astronomy \& Astrophysics, University of California, Santa Cruz, CA 95064, USA}
\affiliation{Department of Astronomy, School of Science, Westlake University, Hangzhou, Zhejiang 310030, China}
\affiliation{Institute for Advanced Study, Tsinghua University, Beijing, 100084, China}

%\correspondingauthor{}
%\email{}

\begin{abstract}
Type II orbital migration is a key process to regulate the mass and semimajor axis distribution of exoplanetary giant planets. The conventional formula of type II migration generally predicts too rapid inward migration to reconcile with the observed pile-up of gas giant beyond 1 au. Analyzing the recent high-resolution hydrodynamical simulations by \citet{Li2024} and \citet{Pan2025} that show robust outward migration of a 
%Jupiter-mass 
gas accreting planet, we here clarify the condition for the outward migration to occur and derive a general semi-analytical formula that can be applied for broad range of planet mass and disk conditions.
The striking outward migration is caused by azimuthal asymmetry in
corotation torque exerted from cicumplanetary disk regions (connecting to horseshoe flow)
%horseshoe flow in the protoplanetary disk 
that is produced by the planetary gas accretion, while the conventional inward migration model is based on radial asymmetry in the torques from the circumstellar protoplanetry disk. We found that the azimuthal asymmetry dominates and the migration is outward, when the gap depth defined by the surface density reduction factor of $1/(1+K')$ is in the range of $0.03 \lesssim K' \lesssim 50$.
Using simple models with the new formula, we demonstrate that the outward migration plays an important role in shaping the mass and semimajor axis distribution of gas giants.
The concurrent dependence of planets' accretion rate and migration direction on their masses
and disk properties
potentially reproduces the observed pile-up of exoplanetary gas giants beyond 1 au, although more detailed planet population synthesis calculations are needed in the future.  
\end{abstract}

\section{Introduction}\label{sec:intro}

Gas giants with masses 
$\gtrsim 0.3\times$Jupiter mass,
%$\gtrsim 0.3M_{\rm J}$ ($M_{\rm J}$ is Jupiter mass), 
which were
discovered by radial velocity (RV) surveys, show
a clear pile-up at %the orbital periods $P \gtrsim 300\, {\rm days}$ or 
the semimajor axes $a \gtrsim 1 \, {\rm au}$.
The occurrence rate of these gas giants jumps up by a factor of more than 3-5 at $1\, \rm au$
%$P \sim 300\, {\rm days}$
and it is almost constant 
%beyond $P \sim 10^3\, {\rm days}$ as long as 
up to $\sim 5\, \rm au$.
%$P \lesssim 10^4\, {\rm days}$
\citep[e.g.,][]{Wittenmyer2020}.
The orbital period bias of RV observations for gas giants would be very weak for $\lesssim 5\, \rm au$, and the possible bias
% and if the possible bias
tends to restrict detection in large $a$ planets. 
Therefore, the pile-up at $a \gtrsim 1 \, {\rm au}$ must be real.

Several papers addressed possible origins of the pile-up.
%using the new type II migration proposed by \citet{Kanagawa2018} and \citet{Robert2018}.
High-resolution hydrodynamical simulations 
found that type II migration of a gas giant is not tied to disk accretion \citep{Duffell2014,Kanagawa2015,DurmannKley2017}, in contrast to  the pioneering studies on gap-opening process \citep[e.g.,][]{Lin1986, Lin1996}.
Accordingly, \citet{Kanagawa2018} proposed that type II migration is no other than
type I migration with lower disk surface density in the bottom of a gap - 
the migration of a high-mass gas giant is slowed down simply by the fraction of surrounding gas it depletes. 
However, the slowing-down due only to the nominal gap formation is not enough to account for the pile-up. 
This is because type I migration is fast and a very deep gap is created only after the planet mass exceeds a Jupiter mass \citep[e.g.,][see Section~\ref{sec:grow_mig}]{Ida2018}.

Type II migration could be further slowed down by additional depletion in the disk surface density
by disk wind \citep{Ida2018}, photoevaporation \citep{Tanaka2020}, or a steep surface density profile \citep{Chen2020b}.
In particular, the global disk gas depletion caused by gas accretion onto the planet
%gas accretion onto a growing gas giant proposed by 
\citep{TanigawaTanaka2016} is very important, which is discussed in detail in Section~\ref{sec:grow_mig}.
Another idea is that planetesimals are initially concentrated in the regions beyond the snow line
\citep[e.g.,][]{Guo2025}.

Migration process itself should also be investigated with high-resolution hydrodynamical simulations with planetary gas accretion. A forming gas giant planet concurrently accretes gas and migrates. Because the accretion modifies the gas surface density in the proximity of the planet, 
gas accretion must be taken into account in migration simulations.

Recently, 
robust outward migration for a Jupiter-mass planet concurrently accreting gas in the disk with the viscosity parameter $\alpha \ga 3 \times 10^{-3}$ was reported by \citet{Li2024} with high-resolution 3D/2D hydrodynamical simulations.
%In general, disk gas cycles in horseshoe orbits many times until it completely crosses the planet orbit. The disk gas from outer regions first follows leading horseshoe orbit, loosing some fraction of gas to the planet, and the gas encounters the planet afterward in the trailing horseshoe orbit with a lower surface density, further losing some fraction of the mass.
They found that this gas loss process results in steady azimuthal asymmetry %between horseshoe flows in 
in the circumplanetary disk (CPD) and horseshoe flows between leading and trailing regions and the asymmetry causes outward migration of the planet.
%, because the surface density of gas inwardly passing in the leading region is always higher than that outwardly passing in the trailing region.
%This mechanism robustly operates as long as most of disk gas from outer region eventually accretes onto the planet.
%As we will show later, a fraction of gas that eventually accretes to the planetary becomes low when $\alpha \la 0.003$ for a Jupiter-mass planet (the mass ratio to the host body is $q=10^{-3})$. 
\citet{Laune2024} also found through a few 2D hydrodynamical simulations, a disk-embedded companion around a host with mass ratio $q = 10^{-3}$ and $\alpha=0.1$ undergoes outward migration if the companion efficiently accretes gas in the accretion disk.   
Their result is consistent with \citet{Li2024},
although they did not explore the dependence on the disk viscosity. 

We note that previous hydrodynamical simulations including the effect of the planetary gas accretion did not find outward migration
\citep{DurmannKley2015,DurmannKley2017,Robert2018}.
\citet{Li2024} argued that 
the outward migration appears as a result of their self-consistent treatments of the planetary accretion and the outer boundary condition to ensure steady accretion of the protoplanetary disk (for boundary conditions, also see \cite{Dempsey2020} and  \citet{Laune2024}), as well as high enough 
resolution for the %circumplanetary disk (CPD).
CPD.

\citet{Pan2025} performed parameter surveys with high-resolution hydrodynamical simulations similar to \citet{Li2024}. They changed planet-to-star mass ratio $q$, the viscosity parameter $\alpha$, and disk aspect ratio $h$ to make clear the parameter ranges for the outward migration.
They found that the relevant parameter range is very wide, as shown in Section~\ref{sec:results}.
These results should impact the arguments on migration of gas giants and the observed pile-up of exoplanetary gas giants at 
$a \gtrsim 1 \, \rm au$.

In this paper, using the results by \citet{Pan2025}, we derive semi-analytical formula for the migration rate as a function of $q$, $\alpha$, and $h$ to show intrinsic physics and the conditions for the outward migration clearly and provide a useful tool for planet population synthesis calculations.
We show that the effects of the azimuthal asymmetry due to  planetary accretion on orbital migration 
(its direction and magnitude) relative to the conventional formula neglecting the azimuthal asymmetry are simply described only by the gap depth parameter \citep{Kanagawa2015}.
%and a parameter of planetary accretion efficiency. 

For the planetary gas accretion rate, we combine the results of \citet{TanigawaTanaka2016}, \citet{Choksi2023}, and \citet{Li2023} to derive a relevant formula.
The thermodynamics (radiation, equation of state) would play a critical role in the accretion rate onto the planet
\citep{AyliffeBate2009,Lambrechts2019,Lega2024} and in migration of accreting Jupiter-mass planets \citep{WuH2024}. 
However, the relevant thermodynamics is not clear yet.
Here, in order to highlight the outward migration by the azimuthal asymmetry created by the planetary gas accretion, we leave the thermodynamic effects for future study.
\citet{Wafflard-Fernandez2025} found outward migration in a giant planet in eccentric orbit in the disk-wind driven accretion disk. 
Significant eccentricity could be excited by planet-disk interaction in the case of high-mass giants \citep[also see][]{Dempsey2021}.
This effect will be commented in this paper, but the details are also left for future studies.

In Section~\ref{sec:method}, we summarize the past results of gap depth created by a planet and gas accretion rate onto the planet. 
Using these formulas and \citet{Pan2025}'s numerical results, in Section~\ref{sec:results},
we derive the outward migration condition and the general migration rate formula of a gas-accreting planet including both outward and inward migrations.

\section{Method}\label{sec:method}

\subsection{Gap depth}
\label{sec:gapdepth}

As a gas giant grows, its gravitational perturbation generates a wide and smooth gap.
The reduction factor of the surface density at the bottom of the gap ($\Sigma_{\rm gap}$) relative to the unperturbed one outside the gap ($\Sigma_{\rm 0}$) is given by
\citep{Kanagawa2015,Dempsey2020}
\begin{equation}
\frac{\Sigma_{\rm gap}}{\Sigma_{\rm 0}} \simeq (1 + K')^{-1} \equiv f_{\rm gap},
\label{eq:gap_0ep}
\end{equation} 
where \footnote{We define $K'=0.04K$ here to distinguish the parameter $K$ usually defined in the literature, where $K = q^2 h^{-5} \alpha^{-1}$.}
\begin{align}
& K' = 0.04 \, q^2 h^{-5} \alpha^{-1} = q_{\rm th} \, q_{\rm vis} \: ; \label{eq:K'}\\
%0.04 \, q_{\rm th}^2 h \, \alpha^{-1}, 
& q_{\rm th} = q /h^{3} = 8 \left(\frac{h}{0.05}\right)^{-3} \left(\frac{q}{10^{-3}}\right);  \label{eq:qth}\\
& q_{\rm vis} = 0.04 \, q_{\rm th} h/\alpha = 16 \left(\frac{h}{0.05}\right)^{-2} \left(\frac{q}{10^{-3}}\right) \left(\frac{\alpha}{10^{-3}}\right)^{-1} ; \\
& q = \frac{m_{\rm p}}{M_*} \: ; \: h = \frac{H}{r},
\label{eq:K}
\end{align}
where $m_{\rm p}$ and $M_*$ are the planet and host star's masses, $H$ is the disk gas scale height and $\alpha$ is the alpha parameter of the turbulent strength.
The conventional thermal condition (the dominance of planetary perturbations over pressure gradient) and viscous gap formation condition
are $q_{\rm th}= 3 \left(r_{\rm H}/H \right)^3 \gtrsim 3$ and
$q_{\rm vis} = 0.04 \, q h^{-2} \, \alpha^{-1} \gtrsim 1.6$, respectively
\footnote{\citet{Lin1986} derived the viscous condition as
$q \gtrsim 40 \nu/r \Omega \simeq 40 \alpha h^2$ $= 1/(0.025 \, h^{-2} \alpha^{-1})$.}. 

% For a Jupiter-mass planet around a sola-mass star ($q \simeq 10^{-3}$) with $h \simeq 0.05$ and $\alpha=10^{-3}$, $q_{\rm th} \simeq 8$ and $q_{\rm vis} \simeq 16$. Accordingly $K' \simeq 130$ and $\Sigma_{\rm gap}/\Sigma_{\rm 0} \simeq 0.008$.  For a Saturn-mass planet ($q \simeq 3\times 10^{-4}$) with $h \simeq 0.06$ and $\alpha=10^{-3}$, $q_{\rm th} \simeq 1.4$ and $q_{\rm vis} \simeq 3.3$. Accordingly $K' \simeq 4.7$ and $\Sigma_{\rm gap}/\Sigma_{\rm 0} \simeq 0.18$. While Jupiter's gap is deep, Saturn's gap is relatively shallow. 

\subsection{The ability of gas accretion onto a giant planet}
\label{sec:gas_acc}

\citet{TanigawaTanaka2016}
semi-analytically predicted the rate of gas accretion onto planet as a function of $m_{\rm p}$.  \citet{rosenthal2020consumption}, \citet{Choksi2023} and \citet{Li2023} revisited this issue by performing high-resolution hydrodynamical simulations.

%We modified the fitting formulas in 
\citet{Choksi2023} and \citet{Li2023} 
performed 2D/3D global simulations with a sink corresponding to planetary gas accretion.
%by adjusting numerical factors, as summarized in the following.
The gas accretion is dominated by Bondi accretion for low-mass planets
and by Hill accretion for hig-mass planets.
The switching mass is $q_{\rm th} \sim 1$, 
because 
\begin{align}
\frac{R_{\rm B}}{H} & = \frac{Gm_{\rm p}}{H \, c_s^2} = q_{\rm th}\\
\frac{R_{\rm H}}{H} & = \left(\frac{q}{3}\right)^{1/3} \frac{r}{H} = 3^{-1/3} q_{\rm th}^{1/3}.
\end{align}
where $R_{\rm B}$ and $R_{\rm H}$ are Bondi and Hill radii, 
and we used $c_s^2 = H^2 \Omega^2 = GM_* h^2 /r$.

If $R_{\rm B} \lesssim R_{\rm H}$, equivalently $q_{\rm th} \lesssim \sqrt{3}/3$, the planet has an envelope rather than a circumplanetary disk.
In this case, $R_{\rm B} \lesssim H$ and the gas accretion onto the planet would be described by Bondi accretion:
\begin{align}
\dot{m}_{\rm B} & \simeq \pi R_{\rm B}^2 \, \rho_{\rm gap} \, c_s = \pi q_{\rm th}^2 \, \rho_{\rm gap} \, H^3 \Omega\label{eq:Bondi_acc}. 
% \\ & = \sqrt{\pi/2} \, q_{\rm th}^2 \, \Sigma_{\rm gap} \, H^2 \Omega, \label{eq:Bondi_acc_Sigma}
\end{align}
Both \citet{Choksi2023} and \citet{Li2023} found the $q_{\rm th}^2$ dependence. 
\citet{Li2023} found a numerical factor several times smaller than $\pi$ in the above and 
%that the results of hydrodynamical simulations (3D with a sink corresponding to planetary gas accretion) are fitted by Eq.~(\ref{eq:Bondi_acc}) with a 7 times smaller factor ($\sim 0.45$ rather than $\pi$). 
%[$\leftarrow$ correct?]}\yp{Yes, it is correct.}
\citet{Choksi2023} found a larger factor.
%that the simulations are fitted with factors $\sim 3.5$ for 3D calculation without a sink and $\sim 12$ for 3D calculation with a sink.
Because they aimed to derive the $q_{\rm th}$ scaling, their inner boundary condition was simple and not consistent with steady accretion with reduction due to the planetary gas accretion (Fig.~{\ref{fig:Sigma_acc} shown later).
Adopting a consistent boundary condition, \citet{Li2024} and \citet{Pan2025} 
obtained a numerical factor consistent with $\pi$, as well as confirmation of the $q_{\rm th}^2$ dependence. 
% \citet{Li2023} pointed out a difficulty to numerically simulate the local density in the gap, $\rho_{\rm gap}$.
Hence, we use Eq.~(\ref{eq:Bondi_acc}).}
For convenience, we use $\Sigma_{\rm gap}$ rather than $\rho_{\rm gap}$ with a simple relation $\Sigma_{\rm gap} = \sqrt{2 \pi} \, \rho_{\rm gap} \, H$ as
\begin{align}
\dot{m}_{\rm B} & = \sqrt{\pi/2} \, q_{\rm th}^2 \, \Sigma_{\rm gap} \, H^2 \Omega.
%, 
\label{eq:Bondi_acc_Sigma}
\end{align}

% Hill radius is given by $R_{\rm H} = (q/3)^{1/3} r = (q_{\rm th}/3)^{1/3} h \, r$.
When $R_{\rm H} \gtrsim H$, equivalently $q_{\rm th} \gtrsim 3$ %(For the transition from Bondi accretion to Hill accretion, see below), 
the gas flow for planetary accretion can be approximated by 2D flow and is supplied by Kepler shear.
In this case,
    \begin{align}
    \dot{m}_{\rm H} & \simeq 2 R_{\rm H} \times \frac{3}{2} R_{\rm H} \Omega \times \Sigma_{\rm gap} \nonumber \\ & = 3\, \left(\frac{q_{\rm th}}{3}\right)^{2/3} \, \Sigma_{\rm gap} 
    \, H^2 \Omega. \label{eq:Hill_acc2}
  %\\ & = 3 \sqrt{2 \pi} \, \rho_{\rm gap} \, (q_{\rm th}/3)^{2/3} H^3 \Omega. \label{eq:Hill_acc}
   \end{align}
All of \citet{Choksi2023}, \citet{Li2023}, and \citet{Pan2025} found the $q_{\rm th}^{2/3}$ dependence.
However, the fitted numerical factor has uncertainty within a factor of 10 for super-Jupiter planets.
\citet{Li2023} and \citet{Choksi2023} found a twice larger factor and a similar factor, respectively.
%$\sim \pi$ rather than $3\sqrt{2\pi}\sim 7.5$ in Eq.~(\ref{eq:Hill_acc}).
%\citet{Choksi2023} found the factor $\sim 9$ fits the simulations both with and without a sink.
\citet{Pan2025} obtained a several times larger factor.
%than $3\sqrt{2\pi}$.
Although \citet{Pan2025} (and \citet{Li2024}) used the inner boundary condition consistent with the steady accretion, they may have had some difficulty in the case of high-mass planet with a very deep gap.
Therefore, in this paper, we use a simple formula given by Eq.~(\ref{eq:Hill_acc2}) for Hill accretion.  
We will discuss the effect of the uncertainty in the numerical factor in Section~\ref{sec:grow_mig}.

In the above, there is a transition regime between the Bondi and the Hill accretion regime $\sqrt{3}/3 \gtrsim q_{\rm th} \gtrsim 3$.
%, while Bondi accretion is applied for $q_{\rm th} $.
%the transition of dominance between $R_{\rm H}$ and $R_{\rm B}$ occurs at $q_{\rm th} = 1/\sqrt{3}$. 
%There is another intermediate regime with $\sqrt{3}/3\lesssim q_{\rm th} \lesssim 3$, which is neglected in this work. Instead, 
We adopt smooth connection between the two regions as Eq.~(\ref{eq:mdotCY}), without discussing the transition regime.
%In this paper, we smoothly connect the Bondi and Hill accretion regimes by comparing the individual accretion rates (Eqs.~(\ref{eq:Bondi_acc}) and (\ref{eq:Hill_acc})), as shown later (Eq.~(\ref{eq:mdotCY})).  

We will use the above \citet{Li2023,Choksi2023} type accretion rate.
On the other hand, we also compare them with \citet{TanigawaTanaka2016}'s formula with in Section~\ref{sec:qth-dependence}.

With the same notations, \citet{TanigawaTanaka2016} derived the semi-analytical formula based on the fitting of 2D local simulations without a sink by \citet{TanigawaWatanabe2002} as

    \begin{align}
    \dot{m}_{\rm BH} & 
    \simeq 0.29 \; q_{\rm th}^{4/3} \Sigma_{\rm gap} \, H^2 \Omega, 
    %\\ & \simeq 0.73 \; q_{\rm th}^{4/3} \rho_{\rm gap} \, H^3 \Omega,
    \end{align}
which we refer as ``TT" formula.    
Although they did not consider 
Bondi accretion regime, 
the dependence $\dot{m}_{\rm BH} \propto q_{\rm th}^{4/3}$ of their formula is in between the Bondi ($\propto q_{\rm th}^{2}$) and Hill ($\propto q_{\rm th}^{2/3}$) scalings. 
As shown below, 
their formula gives approximation of the Li/Choksi scaling \footnote{The 4/3 scaling is also observed in a series of local 3D simulations \citep{Maeda2022,Maeda2024}, though the reasons for its discrepancy with global simulations is unclear.}.

Assuming the steady accretion, the unperturbed disk gas accretion %in inner region 
is
\begin{align}
\dot{m}_{\rm d} & \simeq 3 \pi \Sigma_0 \nu = 3 \pi \alpha \, \Sigma_0 H^2 \Omega. 
% \simeq 3 \pi \sqrt{2 \pi} \, \alpha \, \rho_0 H^3 \Omega.
\label{eq:stellar_mdot}
\end{align}
The combined \textit{local} Bondi and Hill accretion scaled by $\dot{m}_{\rm d}$ is given by
\begin{align}
\frac{\dot{m}_{\rm p,loc}}{\dot{m}_{\rm d}} & = \frac{1}{(1 + K')}\left(\frac{\dot{m}_{\rm d}}{\dot{m}_{\rm B}} + \frac{\dot{m}_{\rm d}}{\dot{m}_{\rm H}}\right)^{-1}, \label{eq:mdotCY}\\
%\dot{m}_{\rm p,loc} & = \frac{1}{1 + K'}\left(\frac{1}{\dot{m}_{\rm B0}} + \frac{1}{\dot{m}_{\rm H0}}\right)^{-1} \\
\frac{\dot{m}_{\rm B}}{\dot{m}_{\rm d}} & = \frac{\sqrt{\pi/2}}{3\pi} \, q_{\rm th}^2 \alpha^{-1} 
%= 133 \, q_{\rm th}^2 \left(\frac{\alpha}{10^{-3}}\right)^{-1},\label{eq:mdotCY2} \\ & 
\ ; \ \frac{\dot{m}_{\rm H}}{\dot{m}_{\rm d}} = \frac{3^{1/3}}{3\pi} \, q_{\rm th}^{2/3} \alpha^{-1}
%= 153 \, q_{\rm th}^{2/3} \left(\frac{\alpha}{10^{-3}}\right)^{-1}, 
\label{eq:mdotCY3} 
\end{align}
and \citet{TanigawaTanaka2016}'s (TT) formula is
\begin{align}
\frac{\dot{m}_{\rm p,loc}}{\dot{m}_{\rm d}} & = \frac{1}{(1 + K')} \times \frac{0.29}{3\pi} \, q_{\rm th}^{4/3} \alpha^{-1}, %= 30.8 \, q_{\rm th}^{4/3} \left(\frac{\alpha}{10^{-3}}\right)^{-1}.
\label{eq:mdotTT2}
% \dot{m}_{\rm p,loc} & = \frac{1}{1 + K'} \, \dot{m}_{\rm BH0} \; ; \; \dot{m}_{\rm BH0} = 0.29 \, q_{\rm th}^{4/3} \Sigma_0 \, H^2 \Omega.
\end{align}
where the subscript ``loc" means the {\it local}  expressions of the planetary gas accretion that does not consider a limit by global disk gas flow and it can exceed $\dot{m}_{\rm d}$.

\subsection{Actual rates of planetary gas accretion and gas flow across the planetary orbit}
\label{sec:qth-dependence}

With $\dot{m}_{\rm p,loc}$ larger than the disk accretion rate $\dot{m}_{\rm d}$, 
the actual efficiency of gas accretion onto a planet, taking account of feedback from the limited supply, is derived as \citep{TanigawaTanaka2016, Rosenthal2020}
 
\begin{align}
f_{\rm p} = \frac{\dot{m}_{\rm p}}{\dot{m}_{\rm d}} & = \frac{\xi}{(1 + \xi)} 
%= \frac{1}{1 + \zeta} 
\: ; \: 
%\zeta = \frac{\dot{m}_{\rm d}}{\dot{m}_{\rm p,loc}},
\xi = \frac{\dot{m}_{\rm p,loc}}{\dot{m}_{\rm d}},
\end{align}
such that when $\xi \gg 1$ (the ability of gas accretion by the planet is large enough), 
this equation shows that most of gas flowing through the disk is accreted onto the planet ($f_{\rm p} = \dot{m}_{\rm p}/\dot{m}_{\rm d} \sim 1$).

\begin{figure*}
\centering
\includegraphics[width=12cm,clip]{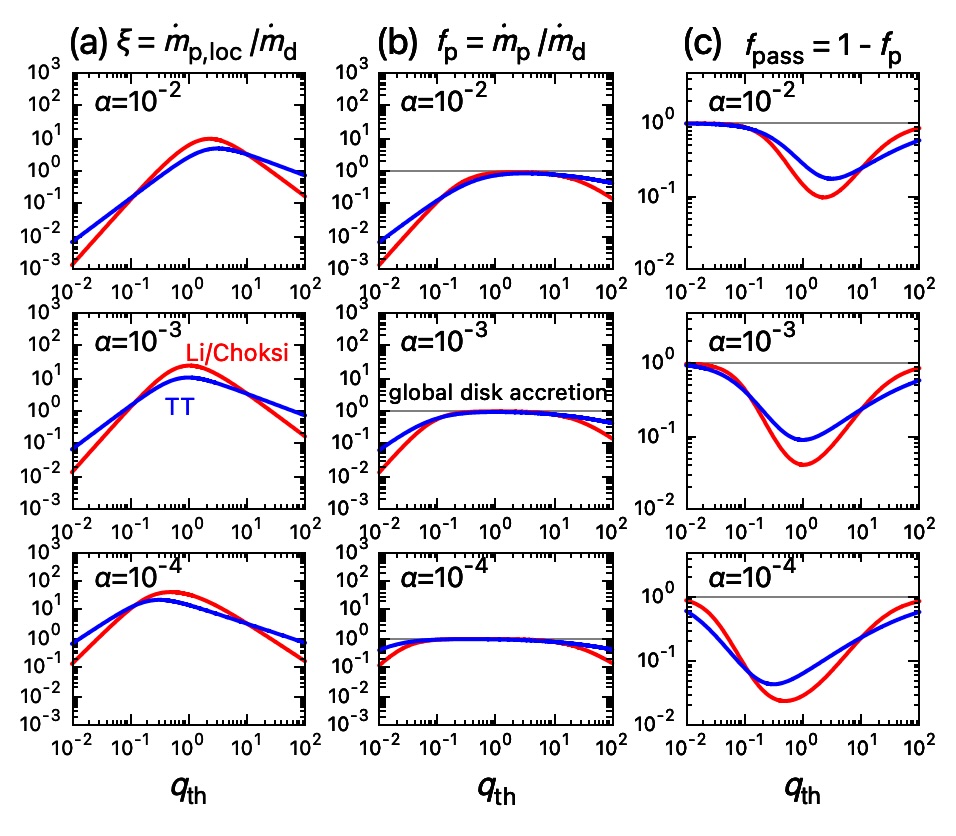}
\caption{The efficiency of gas accretion onto planet as a function of $q_{\rm th}$, predicted by Li/Choksi (red lines) and TT (blue lines) formulas. 
In panels (a), the ability of planetary gas accretion ($\xi = \dot{m}_{\rm p,loc}/\dot{m}_{\rm d}$) are plotted.
The dependence on $\alpha$ is shown in the upper, middle and lower panels, with a fixed value of $h = 0.05$.
In panels (b), the actual ability with a cap of the disk accretion 
($f_{\rm p}=\dot{m}_{\rm p}/\dot{m}_{\rm d}$) is shown.
The panels (c) show the corresponding 
reduction $f_{\rm pass} (=1-f_{\rm p})$ in the global gas surface density.
%due to planetary gas accretion.
}
\label{fig:gas_acc}
\end{figure*}

Figure~\ref{fig:gas_acc}a show
$\dot{m}_{\rm p,loc}/\dot{m}_{\rm d} = \xi$
%1/\zeta$ 
for the Li/Choksi and TT formulas by red and blue lines, respectively.
From Eqs.~(\ref{eq:mdotCY}) and (\ref{eq:mdotCY3}), 
% the parameter $\zeta$ is given explicitly by
\begin{align}
\xi & = \frac{1}{3\pi \alpha (1+K')}
    \left( \sqrt{2/\pi} \, q_{\rm th}^{-2} + 3^{-1/3} q_{\rm th}^{-2/3}\right)^{-1}. 
    % \nonumber\\ & = 3\pi \alpha (1+ 0.04 \, q_{\rm th}^2 h \, \alpha^{-1}) \left( \sqrt{2/\pi}\, q_{\rm th}^{-2} + 3^{-1/3} q_{\rm th}^{-2/3}\right)
%\zeta & = 3\pi \alpha (1+K')\left( \sqrt{2/\pi} \, q_{\rm th}^{-2} + 3^{-1/3} q_{\rm th}^{-2/3}\right) \nonumber\\
    \label{eq:zeta}
\end{align}
In the limits of $q_{\rm th} \gg 1 \ (K' = 0.04 q_{\rm th}^2h/\alpha \gg 1)$, 
%\yp{the power-law index for $h$ seems to be $1/3$?}
\begin{align}
\xi %\simeq 3 \pi \times 0.04 \times 3^{-1/3} \, q_{\rm th}^{4/3} h \nonumber \\
 & \simeq 1.2 \times 10^{2} K'^{-2/3} \,
 h_{0.05}^{-1/3} \, \alpha_3^{-2/3},
%\zeta & \simeq 3 \pi \times 0.04 \times 3^{-1/3} \, q_{\rm th}^{4/3} h \nonumber \\ & \simeq 0.82 \times 10^{-2} K'^{2/3} \, h_{0.05}^{1/3} \, \alpha_3^{2/3}
 % \left(\frac{h}{0.05} \right)^{1/3} \left(\frac{\alpha}{10^{-3}} \right)^{2/3}, 
\label{eq:zeta2}
\end{align}
where 
\begin{align}
 h_{0.05} = h/0.05 \ ; \ \ \alpha_3 = \alpha/10^{-3}.
 \end{align}
 In this case, the gap is so deep that
the accretion onto the planet is limited.

For $q_{\rm th} \ll 1 \ (K' \ll 1)$, 
\begin{align}
%\zeta & \simeq 3\pi \sqrt{2/\pi} \, \alpha \, q_{\rm th}^{-2} \simeq 0.015 \; h_{0.05}/K',
\xi & \simeq (3\pi \sqrt{2/\pi}\, \alpha)^{-1} \, q_{\rm th}^{2} \simeq 66 \; h_{0.05}^{-1} K'.
\label{eq:zeta3}
\end{align}
%where we used $q_{\rm th} = 5 \sqrt{K' \, h^{-1} \alpha}$.
In this limit, the planet gravity is not strong enough to capture most of the disk gas accretion.  
In both limits,
%of $q_{\rm th} \gg 1$ and $q_{\rm th} \ll 1$, 
$\xi \ll 1$ ($\dot{m}_{\rm p,loc} \ll \dot{m}_{\rm d}$).
The effect of the planetary gas accretion is pronounced ($\xi > 1$)
%, equivalently $\dot{m}_{\rm p,loc} > \dot{m}_{\rm d}$) 
for modest values of $q_{\rm th}$.

Figure~\ref{fig:gas_acc}b shows the actual planetary accretion efficiency, $f_{\rm p}=\dot{m}_{\rm p}/\dot{m}_{\rm d} = \xi/(1+\xi)$,
%1/(1+\zeta)$ 
with the Li/Choksi and TT formulas.
The region with $\dot{m}_{\rm p} \sim \dot{m}_{\rm d}$ (full planetary accretion) expands around $q_{\rm th} \sim 1$.

Since the fraction of disk gas accretion that accretes onto the planet is
$f_{\rm p} = \xi/(1 + \xi)$,  
the rest of accretion will pass through the planet's orbit, leading to a reduced disk gas accretion flux and correspondingly, to the reduced surface density $\Sigma_0$ in the inner disk with a factor of
\begin{align}
f_{\rm pass} = \frac{1}{1 + \xi}.
\label{eq:pass}
%\frac{\zeta}{1 + \zeta}.
\end{align}
In both limits of $q_{\rm th} \gg 1$ and $q_{\rm th} \ll 1$, $\xi \ll 1$ and $f_{\rm pass} \simeq 1$.
For modest values of $q_{\rm th}$,
$f_{\rm pass}$ can be significantly smaller than a unity, 
as shown in Figure~\ref{fig:gas_acc}c. The maximum reduction in both paradigms can reach a factor of 10-50 ($f_{\rm pass} \sim 0.02-0.1$). 

We note that this approximation introduces a puzzle 
for massive planets in the limit of $q_{\rm th} \gg 1$
and $\xi \ll 1$.
It is reasonable that $\xi \ll 1$ means $f_{\rm p}\ll 1$, because planetary accretion is quenched by a deepest gap. 
But, the deepest gap should also prevent gas flow through the planetary orbit, which is inconsistent with $f_{\rm pass}\simeq 1$.
We will discuss this issue in Section~\ref{sec:discussion}.

%, where it is shown that $f_{\rm pass}+f_{\rm p}$ is not necessarily $\simeq 1$ but can be $\ll 1$ for $q_{\rm th} \gg 1.$

\subsection{Global reduction in disk gas surface density due to planetary gas accretion}
\label{sec:globaldepletion}

\begin{figure}
\begin{center}
\centering
\includegraphics[width=6.5cm,clip]{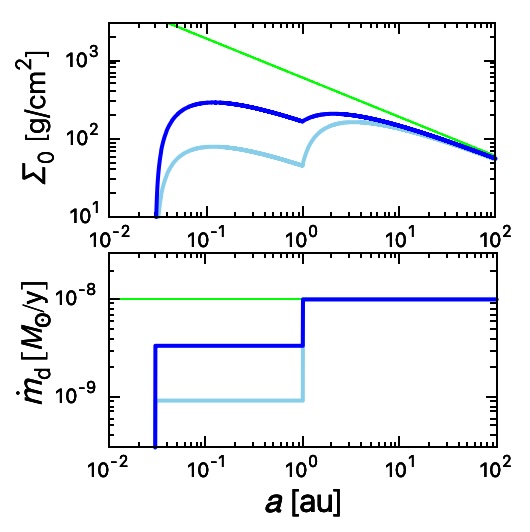}
\caption{The $\Sigma_0$ distribution (the upper panel) and gas accretion rate through the disk (the lower panel) in the steady state with the effect of gas accretion onto the planet and host star (sinks) is plotted in the case of $\dot{m}_{\rm d}=10^{-8} M_\odot/\rm y$, $\alpha = 10^{-3}$, $h = 0.05$, $r_{\rm in}=0.03\, {\rm au}$, and $r_{\rm p}=1\, {\rm au}$.
The green, blue, and skyblue lines are the distribution without any sink, and those with the planet sinks with $\xi=2$ $(f_{\rm p} = 2/3)$ and $\xi=10$ $(f_{\rm p} = 0.91)$, respectively.
For $\alpha = 10^{-3}$ and $h = 0.05$, $\xi=2$ corresponds to $q = 1.6 \times 10^{-5}$ and $q = 1.86 \times 10^{-3}$,
and $\xi=10$ corresponds to $q = 4.2 \times 10^{-5}$ and $q = 4.9 \times 10^{-4}$.
With the sinks, 
$\dot{m}_{\rm d}$ decreases by the accretion onto the planet
inside $1\, {\rm au}$ and vanishes by the accretion onto the host star at $r_{\rm in}$.
\label{fig:Sigma_acc}}
\end{center}
\end{figure}

The gas surface density with the effect of planetary gas accretion under the steady accretion is analytically derived by \citet{TanigawaTanaka2016}:
\begin{align}
    \Sigma_0 = & (1 - f_{\rm p})\left(1-\sqrt{\frac{r_{\rm in}}{r}} \, \right)
    \frac{\dot{m}_{\rm d}}{3 \pi \nu} \hspace{0.5cm} [r_{\rm in} < r < r_{\rm p}], \label{eq:glb_dep1}\\
    \Sigma_0 = &  \left[ (1 - f_{\rm p}) \left(1-\sqrt{\frac{r_{\rm in}}{r}} \, \right) + f_{\rm p} \left(1-\sqrt{\frac{r_{\rm p}}{r}} \, \right)\right] \nonumber\\
    & \times \frac{\dot{m}_{\rm d}}{3 \pi \nu}
    \hspace{0.5cm} [r > r_{\rm p}],\label{eq:glb_dep2}
\end{align}
where $r_{\rm in}$ and $r_{\rm p}$ are the radii of the disk inner edge and the planet's orbit, respectively, and $\nu = \alpha h^2 r^2\Omega$.
Radial profile of $\Sigma_0$ is plotted in Fig.~\ref{fig:Sigma_acc} for $\xi=2$ $(f_{\rm p} = 2/3)$ and $\xi=10$ $(f_{\rm p} = 0.91)$.

This analytical formula for $\Sigma_0$ distribution with a planet sink was reproduced by 3D global hydrodynamical simulation (Figure~D2 in \citet{Li2023}).
Thus, the unperturbed (no gap) disk density at $r\simeq r_{\rm p}$ is reduced by a factor of $(1 - f_{\rm p})$ from that at large distance to the planet's orbit.
%\footnote{\rev{In these simulations, planet's $r_{\rm p}$ is fixed such that its migration does not contribute to the total (disk+planet) mass flux (see Section~\ref{sec:discussion}.}}

%One can alternatively calculate $f_{\rm pass}$ based on the 4/3 scaling of \citet{TanigawaTanaka2016}.

The actual surface density is further reduced by gap opening from this value of $\Sigma_0$. In the remainder of this paper, for more direct comparison with the results from \citet{Li2024} and \citet{Pan2025}, we adopt the accretion rate scalings in Eqs.~(\ref{eq:mdotCY}) and (\ref{eq:mdotCY3}) in calculating the planet accretion rate $f_{\rm p}$
%unless stated otherwise.

If $f_{\rm p}$ is large enough, say $\xi \gtrsim 2$, 
the planetary accretion effect in the disk global depletion is significant (see Fig.~\ref{fig:Sigma_acc}). 
The condition 
%$f_{\rm pass} \lesssim 0.5$ is equivalent to  
$\xi \gtrsim \xi_{\rm acc} = 2$ reads as (Eq.~(\ref{eq:zeta3})
with $q_{\rm th} \ll 1$)
% \begin{align} 0.5 \left(\frac{\zeta_{\rm acc}}{0.5}\right) & \gtrsim 3\pi \sqrt{2/\pi} \, \alpha \, q_{\rm th}^{-2} \simeq 0.015 \left(\frac{h}{0.05} \right) \frac{1}{K'}.
%\simeq 0.0075 \left(\frac{\alpha}{10^{-3}} \right) q_{\rm th}^{-2}, \end{align} and this equation is equivalent to
\begin{align}
K' \gtrsim K'_{\rm acc} \equiv \,  0.03 \, h_{0.05} \left(\frac{\xi_{\rm acc}}{2}\right).
% q_{\rm th} & \gtrsim 0.19 \left(\frac{\zeta_{\rm acc}}{0.1}\right)^{-1/2} \left(\frac{\alpha}{10^{-3}}\right)^{1/2}.
\label{eq:qthlow}
\end{align}
With $K' = 128\, h_{0.05}^{-5} (q/10^{-3})^2 \alpha_3^{-1}$ (Eq.~(\ref{eq:K'})), Eq.~(\ref{eq:qthlow}) is
\begin{align}
q \gtrsim 1.5 \times 10^{-5} \, h_{0.05}^{3} \, \alpha_3^{1/2}\left(\frac{\xi_{\rm acc}}{2}\right)^{1/2}. 
\end{align}
We point out that even a super-earth inside a very shallow gap ($0.03 \lesssim K' \ll 1$) 
accretes most of disk gas flow through it.  However, Eq.~(\ref{eq:zeta}) assumes that
the planet has already bypassed the Kelvin-Helmholtz contraction limit and is undergoing
runaway gas accretion.
Equivalently, $m_{\rm p}$ must be larger than the critical core mass $\sim 10 M_\oplus$ ($q \sim 3 \times 10^{-5}$).  

\subsection{Conventional type-I and type-II migration rates}\label{sec:mig_rate}

The conventional type-I migration rate is given by \citet{Tanaka2002}
\begin{align}
\frac{\dot{a}_{\rm I}}{a}  \sim 
- \, q \; \frac{\pi \, \Sigma_0 \, r_{\rm p}^2}{M_*} h^{-2} \Omega,
\label{eq:typeI}
\end{align}
where $\Sigma_0$ is an unperturbed local surface density of disk gas.
In this paper, we assume the planet orbit is circular and the orbital radius $r_{\rm p}$ is equal to the semimajor axis $a$ of the planet. 
%In this paper, 
Here, we define {\it outward} migration rate with a plus sign and {\it inward} disk accretion rate $\dot{m}_{\rm d}$ 
%\begin{align} \dot{m}_{\rm d} = 3 \pi \, \Sigma_0 \, \alpha h^2 r^2 \Omega. \label{eq:disk_steady_acc} \end{align} In this paper, the disk accretion rate 
is defined with a plus sign,
%if it is inward, 
in order to make clear the planet migration direction.
Using the steady disk accretion rate 
$\dot{m}_{\rm d} = 3 \pi \, \Sigma_0 \, \alpha h^2 r^2 \Omega,$
%Eq.~(\ref{eq:disk_steady_acc}),
Eq.~(\ref{eq:typeI}) is
\begin{align}
\frac{\dot{a}_{\rm I}}{a} \sim -\frac{q}{3\alpha h^4} \frac{\dot{m}_{\rm d}}{M_*}
% = - \frac{5}{3} K'^{1/2} h^{-3/2}\alpha^{-1/2}\frac{\dot{m}_{\rm d}}{M_*},
\simeq - 4.7 \times 10^3 K'^{1/2}
\alpha_3^{-1/2}
h_{0.05}^{-3/2} \frac{\dot{m}_{\rm d}}{M_*}.
% = q_{\rm th} \; \frac{1}{3\alpha h}\frac{\dot{m}}{M_*}.
\label{eq:typeI2}
\end{align}
where we used $q = 5 \sqrt{ K' h^{5} \alpha}$ 
(Eq.~\ref{eq:K'}).
Because the effect of gas accretion onto a planet on migration
is regulated by $K'$, as shown below, we use type I migration formula 
as a function of $K'$ rather than $q$.  

According to \citet{Kanagawa2018}, type II migration rate is fitted by
\begin{align}
\frac{\dot{a}_{\rm II,gap}}{a} \simeq \frac{1}{1+K'}\, \frac{\dot{a}_{\rm I}}{a},
\label{eq:typeII}
\end{align}
which we will refer to as the %``Kanagawa" formula or
``type II (gap)" formula.
\citet{TanigawaTanaka2016} and \citet{Tanaka2020} proposed
the further reduction of type II migration rate caused by the
global gas surface density depletion due to gas accretion onto a planet (Fig.~\ref{fig:Sigma_acc}).
% According to Eqs.~(\ref{eq:glb_dep1}) and (\ref{eq:glb_dep2}), in the limit of large $q_{\rm th}$ iand small $f_{\rm p}$, this reduction only affects $r < r_[\rm p}$.  Consequently, the classical type II migration due to the Lindblad torque at $r>r_{\rm p}$ still exists even when the gap is extremely clean in the large $q_{\rm th}$ limit \citep{Kanagawa2017, Chen2023}.
For a modest value of $q_{\rm th}$ ($\sim 1$), the global surface density reduction significantly slows down migration as well as the local surface density reduction by gap opening.
Accordingly, the type II migration rate is modified to be 
\begin{align}
\frac{\dot{a}_{\rm II,gap+dep}}{a} \sim \frac{f_{\rm pass}}{(1+K')}\, \frac{\dot{a}_{\rm I}}{a},
\label{eq:typeII2}
\end{align}
which we will refer to as the
%``Tanigawa-Tanaka" formula or 
``type II (gap+dep)" formula.
%For $q_{\rm th} \ll 1$, $K' \ll 1$ and \begin{align}\frac{\dot{a}_{\rm II}}{a} & \sim \frac{\dot{a}_{\rm I}}{a} %\nonumber \\ & \simeq - 4.7 \times 10^3 K'^{1/2}\alpha_3^{-1/2}h_{0.05}^{-3/2} \frac{\dot{m}_{\rm d}}{M_*}.\label{eq:typeI3}\end{align}
%where $h_{0.05} = h/0.05$ and $ \alpha_3 = \alpha/10^{-3}$.
%Since $K' \gg 1$ for $q_{\rm th} \gg 1$, we obtain\begin{align}\frac{\dot{a}_{\rm II}}{a} & \sim \frac{f_{\rm pass}}{K'}\, \frac{\dot{a}_{\rm I}}{a} \nonumber \\ & \sim - \, 4.7 \times 10^3 f_{\rm pass} \, K'^{-1/2}\alpha_3^{-1/2}h_{0.05}^{-3/2} \frac{\dot{m}_{\rm d}}{M_*}.\label{eq:typeII4}\end{align}

Because $K'$ and $f_{\rm pass}$ (equivalently, $\xi$) are functions of only $\alpha, h$ and $q$, the theoretical migration rates scaled by $\dot{m}_{\rm d}/M_*$
are also functions of only $\alpha, h$ and $q$, independent of 
%the disk surface density
$\Sigma_0$, as long as steady accretion is established.

\section{Results}\label{sec:results}

\subsection{Summary of hydrodynamical simulations}

We first summarize the high-resolution hydrodynamical simulations by \citet{Li2024} and \citet{Pan2025}.
They showed that gradual mass loss from the disk gas in horseshoe orbits to the planet through a circumplanetary disk (CPD) during encounters in many horseshoe excursions starting in the leading side, results in slightly higher gas surface density in the leading CPD region than in the trailing region.
Here we define ``CPD" as the dense region rotating around the planet in prograde direction that is connected to horseshoe flow.
In an Eulerian view, 
the steadily higher surface density in leading side
% is higher than that in trailing side, the disk gas 
always accelerates the planet orbital motion, resulting in positive net angular momentum transfer from the disk to the planet. 
In other words, corotation torque that is azimuthally asymmetric between the leading and trailing sides of the planet 
changes migration speed and even direction in some parameter range. 

\begin{figure*}[htbp]
\begin{center}
\includegraphics[width=160mm,angle=0]{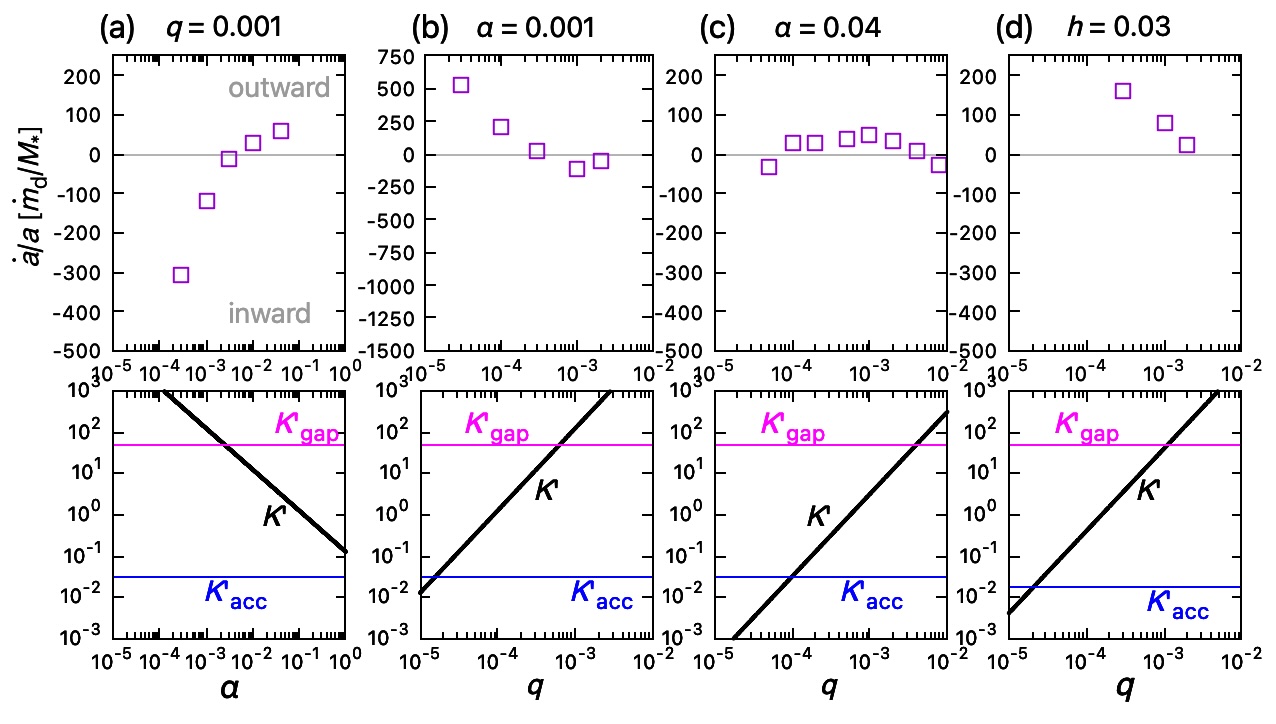}
\caption{ 
The migration rates obtained by \citet{Pan2025} and \citet{Li2024} (upper panels).
The scaling factor is $\dot{m}_{\rm d}/M_*$, where $\dot{m}_{\rm d}$ is the disk mass accretion rate and $M_*$ is the host star mass. The simulation parameters are the planet to star mass ratio $q=m_{\rm p}/M_*$, the disk viscosity parameter $\alpha$, and the disk aspect ratio $h = H/r$.
Except for panel (d), $h = 0.05$ is used.
In panel (d), $\alpha = 0.04$.
The vertical scales are different in panel (b) for visual convenience. 
In each of the lower panels, $K'$ corresponds to the parameters in the directly upper panel.
The outward migration is predicted 
%by the region of 
in $K'_{\rm acc} < K' < K'_{\rm gap}$
for $\xi_{\rm acc} = 2$ ($K'_{\rm acc} = 0.03$) and $K'_{\rm gap}=50$.
\label{fig:mig}}
\end{center}
\end{figure*}

\citet{Pan2025} performed systematic surveys of simulations to show how the migration rate 
depends on the simulation parameters, $\alpha$, $q$, and $h$.
%The disk aspect ratio is $h = 0.05$ except $h = 0.03$ in panel \mg{d}.} 
%changes as one simulation parameter from $\alpha, h$ and $q$ changes with other parameters fixed.
The upper panels of Figure~\ref{fig:mig} summarize their surveys
(square symbols) and show that migration is outward ($\dot{a}/a > 0$)
in wide ranges of parameters.

\citet{Li2024} already found that outward migration of a Jupiter-mass planet ($q = 10^{-3}$) changes to inward migration for $\alpha \lesssim 3 \times 10^{-3}$, which is shown in
panel (a).
In panel (b), for a fixed value of $\alpha$ $(= 10^{-3})$,
the $q$ dependence is shown.
Outward migration is found in lower $q$ region.
These two figures suggest that outward migration is not favored in too deep gap cases ($K'\propto q^2 \alpha^{-1} h^{-5}$).

%, and it gradually reduced to type II (gap) formula as $q$ increases.
Panel (c) shows the $q$ dependence with higher $\alpha$ $(=0.04)$
than that in Panel (b).  
The outward migration region expands to higher $q$ region here.
In Panel (d), while $\alpha=0.04$ is the same as Panel (c), smaller $h$ ($=0.03$) is used. 
The outward migration region shifts lower $q$ region than in Panel (c).
These two figures strongly support the above idea of gap depth.
 
On the other hand, Panel (c) shows that there is also a lower limit in gap depth for outward migration to occur.
Based on these hydrodynamical simulation survey results,
we next derive the condition for outward migration. 
%The panel c with $\alpha = 0,04$ ($h = 0.05$) and the panel d with $\alpha = 10^{-3}$ and $h = 0.03$ show a qualitatively similar behavior that there is an upper limit in $q$ for outward migration. However, the panel c shows that there is a lower limit too.

\subsection{Outward migration conditions}
\label{sec:criteriaoutward}

Analyzing the summarized simulation results in upper panels of Fig.~\ref{fig:mig} in details,
we find that the migration is outward, if the gap is not too deep to maintain the dominance of horseshoe corotation torque over Lindblad torque (detailed discussion is given in Section~\ref{sec:general})
and if the planet accretes most of disk gas flow to retain the azimuthal horseshoe asymmetry. 
Therefore, the gap-depth parameter $K'$ [$\Sigma_{\rm gap}/\Sigma_0 = 1/(1+K')]$ is the most important parameter for the occurrence of the outward migration. 

In the lower panels of Fig.~\ref{fig:mig}, we plot $K'$ analytically determined from $\alpha, h$ and $q$ in each of the parameter surveys as 
$K' = 0.04 \, q_{\rm th}^2 h \, \alpha^{-1} = 2 \, h_{0.05} \, \alpha_3^{-1} q_{\rm th}^2$ where $q_{\rm th} = 8 \, h_{0.05}^{-3} (q/10^{-3})$.
%In each set of surveys (each panel in Fig.~\ref{fig:mig}),  the outward migration is found in broad range of parameters. 

We find that the outward migration regime corresponds to similar range of $K'$ in all the panels (all of the parameter surveys).
The dependence on either of $\alpha, h$ and $q$ is reduced to that on the single parameter $K'$.

%This effect is quantitatively evaluated by the difference from the above analytical estimate, \rev{as shown in Section~\ref{sec:results}.

The upper limit of $K'$ for the outward migration is found as 
\begin{align}
K' \sim K'_{\rm gap} = 50.
\end{align}
\citet{Li2024} found that outward migration of a planet with $q = 10^{-3}$ is turned to inward migration for $\alpha \lesssim 3 \times 10^{-3}$ (Fig.~\ref{fig:mig}a).  Below this critical $\alpha$ value, the
gap is so deep that the horseshoe corotation torque no longer dominates the Lindblad torque, and the gas accretion onto the planet, which makes the azimuthal asymmetry, is quenched.
%(also see Section~\ref{sec:discussion} for possible breakdown of steady disk gas accretion in high $K'$ cases).} 
%{\color {red} But would there be any gas flow through the gap for $K^\prime \gtrsim K^\prime _{\rm gap}$?  If not, would the steady state mass flux assumption break down?}

On the other hand, the lower limit of $K'$ for the outward migration
is $K'_{\rm acc} \sim 0.03$, as shown in Fig.~\ref{fig:mig}. 
From Eq.~(\ref{eq:zeta3}) for $q_{\rm th} \ll 1$,
the upper limit of $\xi$ $(>1)$ for the outward migration is
\begin{align}
%\zeta_{\rm acc} \simeq 0.5 \, h_{0.05} \, \left( \frac{ K'_{\rm acc}}{0.03} \right)^{-1}.
\xi_{\rm acc} \simeq 2 \, h_{0.05}^{-1} \, \left( \frac{ K'_{\rm acc}}{0.03} \right).
\end{align}
The fraction of accretion onto a planet in upstream gas accretion through the disk is 
given by $\xi/(1+\xi)$, which is $2/3$ for $\xi = 2$,
%$1/(1+\zeta)$, which is $2/3$ for $\zeta = 0.5$,
If more than two thirds of disk gas accretes onto a planet, 
the resultant azimuthally asymmetric torque reverses the migration direction.
For the parameters, $q = 10^{-3}$, $\alpha=0.1$, and $h = 0.07$, where \citet{Laune2024} found outward migration, $K' \simeq 0.24$, which is also within the above outward migration range.

% Using Eq.~(\ref{eq:K'}), the former is rewritten as 
% \begin{align} \alpha & \gtrsim 0.04 \, q_{\rm th}^2 h /K'_{\rm gap}  = 5 \times 10^{-5} \left(\frac{K'_{\rm gap}}{50}\right)^{-1}\left(\frac{h}{0.05}\right) q_{\rm th}^2, \label{eq:outmig_a1} \\ {\rm or} &  \nonumber \\ q_{\rm th} & \lesssim 4.5 \left(\frac{K'_{\rm gap}}{50}\right)^{1/2} \left(\frac{h}{0.05}\right)^{-1/2} \left(\frac{\alpha}{10^{-3}}\right)^{1/2}.\label{eq:outmig_q1} \end{align}
% and (\ref{eq:outmig_q1}), the condition for outward migration is 
% \begin{align} 0.19 \left(\frac{\zeta}{0.5}\right)^{-1/2} & \lesssim \frac{ q_{\rm th} }{(\alpha/10^{-3})^{1/2}} \lesssim 4.5 \left(\frac{K'_{\rm gap}}{50}\right)^{1/2} \left(\frac{h}{0.05}\right)^{-1/2} \label{eq:outmig_q2} \end{align} Equivalently, this condition is 
% \begin{align} 4.9 \times 10^{-5} \left(\frac{K'_{\rm gap}}{50}\right)^{-1} \left(\frac{h}{0.05}\right)  & \lesssim \frac{ \alpha}{q_{\rm th}^2} \lesssim 2.8 \times 10^{-2}\left(\frac{\zeta}{0.5}\right) \label{eq:outmig_a2} \end{align}
Thus, the condition for outward migration is simply given by
$K'_{\rm acc} \lesssim K' \lesssim K'_{\rm gap}$, or more explicitly by
\begin{align}
0.03 \, h_{0.05} \left( \frac{\xi_{\rm acc}}{2}\right) 
%\left( \frac{\zeta_{\rm acc}}{0.5}\right)^{-1} 
& \lesssim K' \lesssim 50 \left(\frac{K'_{\rm gap}}{50}\right). 
\label{eq:outward_K'}
\end{align}
The parameters $K'_{\rm gap}$ $(\gg 1)$ and $\xi_{\rm acc}$ $(>1)$ should be empirically determined
by comparison with the hydrodynamical simulation results, such as
$K'_{\rm gap} \sim 50$ and $\xi_{\rm acc} \sim 2$.
The range of outward migration, $0.03 \lesssim K' \lesssim 50$, is broad enough to impact planet formation models.
In the case of $\alpha = 10^{-3}$ and $h = 0.05$, the outward migration range is as broad as $q = 1.5 \times 10^{-5}$ to $q = 6.3 \times 10^{-4}$ (from super-earths to 2 Saturn's mass gas giants).   
We note that $q$ must be larger than the critical core, $q \sim 3 \times 10^{-5}$,  
to initiate the runaway gas accretion and generate the asymmetry by the accretion onto the planet.
It is safer to set the lower limit at $q \sim 3 \times 10^{-5}$.

Because the outward migration condition is determined solely by $K'$,
we can easily derive a (semi-) analytical formula of migration rate including outward migration
%due to the azimutial asymmetry caused by planetary gas accretion as we will show 
as in Section~\ref{sec:general}.

The semi-analytical formula including the azimuthal asymmetry in addition to the effects of the gap opening and the global depletion will be referred to as ``type II (full)" formula.

\subsection{General semi-analytical formula for orbital migration rates including both outward and inward migations}
\label{sec:general}

\begin{figure*}[htbp]
\begin{center}
\includegraphics[width=160mm,angle=0]{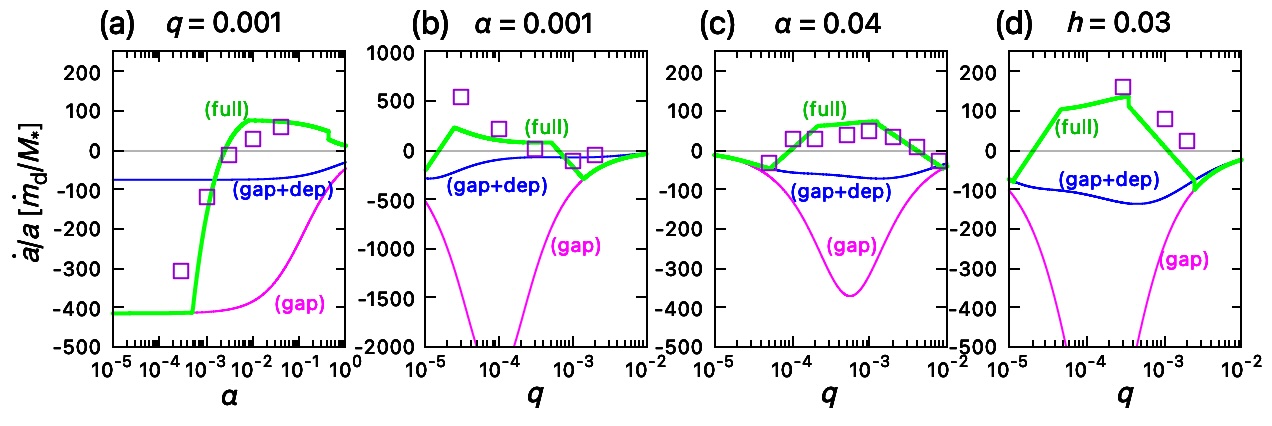}
\caption{\label{fig:mig_anly} 
The general semi-analytical migration formula [type II (full) formula]
including the azimuthal asymmetry in addition to the effects of the gap opening and the global depletion due to the planetary gas accretion (green lines).
It is compared with 
the type II (gap) formula with only the gap opening effect (magenta lines),
the type II (gap+dep) formula with the gap opening and the global depletion effects (blue lines), and
four systematic parameters survey results of hydrodynamical simulations by \citet{Pan2025} (brown squares), which are the same panels as in the upper panels in Fig.~\ref{fig:mig}. 
}
\end{center}
\end{figure*}

The results of \citet{Pan2025}'s simulations are compared with the existing type II (gap) and type II (gap+dep) formulas in Fig.~\ref{fig:mig_anly}.
It shows that 
% the migration rate with the effect of gas accretion onto the planet in the outward migration region is fitted in broad ranges of parameters $q$, $\alpha$, and $h$, by 
%\yp{the sign below for $\dot{a}$ is positive, and the classical migration should be negative, right? I find that the above formula for the classical migration rate does not have a minus sign.}
\begin{align} 
 \frac{\dot{a}}{a} & \simeq - \left( \frac{\dot{a}}{a} \right)_{\rm gap+dep} 
   \ \ \ \ \ \ [K'_{\rm acc} \lesssim K' \lesssim K'_{\rm gap}],
   \label{eq:outmigrate}
\end{align} 
where $K'_{\rm acc}\sim 0.03$ and $K'_{\rm gap}\sim 50$.
Because $(\dot{a}/{a})_{\rm gap+dep}$ in Eq.~(\ref{eq:outmigrate}) is negative (inward), the fitting rate, $\dot{a}/{a}$, is positive (outward).
The type II (gap+dep) formula is given by Eq.~(\ref{eq:typeII2}) with Eq.~(\ref{eq:typeI2}):
\begin{align}
& \left(\frac{\dot{a}_{\rm II}}{a}\right)_{\rm gap+dep} 
 \simeq \frac{1}{(1+K')} \, 
 %\frac{\zeta}{1+\zeta} \,
 \frac{1}{(1+\xi)} \,
 \frac{\dot{a}_{\rm I}}{a}, \label{eq:adot_a2} 
\end{align}
where $\dot{a}_{\rm I}/{a}$ is given by Eq.~(\ref{eq:typeI2})
\begin{align}
& \frac{\dot{a}_{\rm I}}{a}\simeq - 4.7 \times 10^3 K'^{1/2}
\alpha_3^{-1/2}
h_{0.05}^{-3/2} \frac{\dot{m}_{\rm d}}{M_*}, \label{eq:adot_a1} 
% & \left(\frac{\dot{a}_{\rm II}}{a} \right)_{\rm gap} \simeq \frac{1}{1+K'}\, \frac{\dot{a}_{\rm I}}{a} \\
\end{align}
and 
% with $q_{\rm th} = 5 \sqrt{K' \, h^{-1} \alpha}$ (Eq.~\ref{eq:K'}), 
$\xi$ is given as a function of $K', \alpha$ and $h$ by Eq.~(\ref{eq:zeta}) with Eqs.~(\ref{eq:K'}) and (\ref{eq:qth}).
%\begin{align}
%& \zeta = 3\pi \alpha (1+K') \left( \sqrt{2/\pi} \, q_{\rm th}^{-2} + 3^{-1/3} q_{\rm th}^{-2/3}\right), \\
%& \xi = (3\pi \alpha (1+K'))^{-1} \left( \sqrt{2/\pi} \, q_{\rm th}^{-2} + 3^{-1/3} q_{\rm th}^{-2/3}\right)^{-1}, \label{eq:formula} \end{align}

A possible interpretation of Eq.~(\ref{eq:outmigrate}) is as follows.
\citet{TanakaOkada2024} analytically derived 3D Lindblad and corotation torques due to radial asymmetries in the protoplanetary disks: the torque components due to curvature, the radial gradients of $\Sigma_0 \propto r^{-\zeta}$, $P \propto r^{-\delta}$,
$H \propto r^{\mu}$, and $T \propto r^{-\chi}$.
Assuming steady disk gas accretion $\mu = -\chi/2 + 3/2$ ($3 \pi \Sigma_0 \nu = 3 \pi \Sigma_0 \alpha H^2 \Omega$ is constant of $r$), ideal gas $\delta = \zeta + \mu + \chi$ ($P \propto \rho T \propto \Sigma_0 T / H$),
and that the local gas surface density is uniform, they derived the Lindblad and corotation torques as
\begin{align}
\Gamma_{\rm L}/\Gamma_0 & = -(2.382 + 0.094 \, \zeta + 1.297 \, \chi) \label{eq:Lind} \\ 
\Gamma_{\rm c}/\Gamma_0 & = 0.946 - 0.630 \, \zeta + 0.858 \, \chi, \label{eq:corot}
\end{align}
where $\Gamma_0 = q^2 h^{-2} \Sigma_0 r^4\Omega^{2}$.
In the no gap cases, 
the total torque\footnote{For $\zeta = 0.5$ and $\chi=1$ that \citet{Pan2025} adopted, $\dot{a}/a = 2\, \Gamma \simeq -4.5 \Gamma_0$ , which is consistent with Eq.~(\ref{eq:typeI}).} $\Gamma = \Gamma_{\rm L} + \Gamma_{\rm c} = - (1.436 + 0.724 \, \zeta + 0.439 \, \chi) \, \Gamma_0$ is always negative (inward migration) for $\zeta, \chi > 0$.  
Both $\Gamma_{\rm L}$ and $\Gamma_{\rm c}$ are included
in the calculation of ${\dot a}_{\rm I}$.

In the gap opened cases, $\Sigma$ locally (in radial scale $\ll r$) changes. 
Because the gap half width is relativey large (Eq.~(\ref{eq:gap_width})), Lindblad resonances, which are broadly distributed in the gap bottom regions, are significantly weakened.
On the other hand, a CPD is formed when a gap is opened. 
Because the scale height of the CPD is smaller than that of the circumstellar (protoplanetary) disk \citep{Li2023}, the density is higher in the CPD. 
Accordingly, corotation torque exerted from the horseshoe flows connected to the CPD  dominates Lindblad torque and 
$\Gamma \sim \Gamma_{\rm c}$, in the case of a modest gap.

%and the local value of $|\alpha|$ would be $\ll 1$, because the local gap shape would be radially symmetric, 
% In the circumplanetary environment, the effective local $\alpha$ may be small \citep{Fujii2014}. 
Furthermore, the azimuthal asymmetry enhances $\Gamma_{\rm c}$. 
As we already described, the disk gas repeats many encounters during many horseshoe excursions, gradually loosing mass to the planet,
until it completely passes the planet orbit. 
Because the mass loss starts in the leading horseshoe region, 
slightly higher gas surface density is established in the leading horseshoe region than in the trailing region and this azimuthal asymmetry causes outward migration.
\citet{Pan2025} (see also \cite{Li2024}) shows that the migration is outward at $q_{\rm th} \sim O(1)$ if the planetary accretion is included, while it is still inward otherwise (the leading and trailing flows are symmetric).

The degree of the azimutial asymmetry is estimated as follows.
As discussed in Section~\ref{sec:gas_acc}, the rate of planetary gas accretion is self-regulated to be equal to $\dot{m}_{\rm d}$
%the gas accretion through the disk,
when outward migration occurs.
The gas in horseshoe excursions would drifts inward on average by the disk global viscous diffusion.   
The timescale for the mean disk gas flow to cross the gap by the radial drift velocity due to the diffusion ($v_r \simeq -(3/2) \nu/r$)
%\simeq -(3/2) \alpha H^2 \Omega/r$) 
is 
\begin{align}
t_{\rm cross} \sim \frac{2\Delta_{\rm gap}}{3 \alpha H^2 \Omega/2r} 
= \frac{4 (\Delta_{\rm gap}/r)}{3 \alpha h^2\Omega},
\end{align}
where $\Delta_{\rm gap}$ is the gap half-width.
The excursion timescale of a horseshoe orbit is
\begin{align}
t_{\rm hs} \sim \frac{2\pi r}{(3/2) \Delta_{\rm gap}\Omega}
= \frac{4\pi}{3 (\Delta_{\rm gap}/r)\Omega}.
\end{align}
Because the horseshoe flow loses almost all of the mass in $t_{\rm cross}$, the averaged decrease in $\Sigma$ due to the planetary gas accretion during one horseshoe encounter is $\Delta \Sigma \sim \Sigma/N$, where 
%$N$ is horseshoe excursions occurring during $t_{\rm cross}$ and it is given by 
$N \sim t_{\rm cross}/t_{\rm hs}$.
Consequently, the azimuthal asymmetry  
%relative difference in the surface density 
between the leading and trailing horseshoe regions would be 
\begin{align}
\frac{\Delta \Sigma}{\Sigma} 
\sim \frac{t_{\rm hs}}{t_{\rm cross}} \sim \pi \frac{\alpha h^2}{(\Delta_{\rm gap}/r)^2}.
\label{eq:DSig}
\end{align}
\citet{Kanagawaetal2015} derived an empirical formula for $\Delta_{\rm gap}$ by the definition of $\Sigma = \Sigma_0/2$. 
Because we are concerned with the edge of the gap bottom, we use a half value of their $\Delta_{\rm gap}$ as
%derived by \citet{Kanagawaetal2015}, 
\begin{align}
\frac{\Delta_{\rm gap}}{r} & = \frac{1}{2} \times \left[ 0.41  \left(\frac{K' h^2}{0.04}\right)^{1/4} \right].
\label{eq:gap_width}
\end{align}
Substituting this equation into Eq.~(\ref{eq:DSig}), we obtain
\begin{align}
\frac{\Delta \Sigma}{\Sigma} 
\sim  \frac{(4/5)\pi}{0.41^2} K'^2 \alpha h \sim 1.5  \left(\frac{K'}{10}\right)^2 \alpha_3 \, h.
\label{eq:DSig2}
\end{align}
%where we assume $K'\sim 1.22$ as a logarithmic mean value of $K'$ for outward migration ($0.03 \, h_{0.05} \lesssim K' \lesssim 50$).

%\yp{may be replaced the timescale $t_{\rm hs}$ or $t_{\rm cross}$ with the corresponding velocity $v_{\rm hs}$ or $v_{\rm cross}$? }
%\yp{what is the role of accretion here?} 
Therefore, the azimuthal torque asymmetry would be $\sim h$, and it is comparable to the radial torque asymmetry \citep[e.g.,][]{Miyoshi1999,KleyNelson2012}.
This explains why the absolute value of outward migration rate due to the azimuthal asymmetry is comparable (at least by the order of unity) to that of type II (gap+dep)  inward migration rate due to the radial asymmetry (Eq.~\ref{eq:outmigrate}).

%Because the gas surface density difference due to the planetary accretion between the leading and trailing regions is relatively small, the contribution to the torque from the azimuthal asymmetry may be the order of that from the radial asymmetry. These interpretations may explain the torque with the opposite sign and the same order of absolute value as the Tanigawa-Tanaka formula (``type II (gap+dep)" in Fig.~1).   

In order to use Eq.~(\ref{eq:outmigrate}) for planet population synthesis, we need to derive the fitting formula including the transition regimes between the outward and ordinary inward migration regimes
%and the type II (gap+dep) inward migration regime, 
at least continuously (we do not require smoothness).  
Here we apply a log-linear  interpolation of $K'$ on the $\dot{a}/a-q$ or $\dot{a}/a-\alpha$ plots in Fig.~\ref{fig:mig_anly}.
We define detaching points as the linear interpolation meets type II (gap+dep) formula in these plots.
We empirically find that the hydrodynamical simulation results in Fig.~\ref{fig:mig_anly} meet the interpolation at
the higher parameter $K' \sim K'_{\rm gap,h} = 5\, K'_{\rm gap} = 250$
and at the lower one $K' \sim K'_{\rm acc,l} 10^{-1/2} = K'_{\rm acc} = 0.01 \,h_{0.05}$. %\yp{may need to define where are the detaching points? }
We log-linearly connect the fitting migration rate given by
Eq.~(\ref{eq:outmigrate}) to 
$(\dot{a}/a)_{\rm gap+dep}$ with the condition of 
$\dot{a}/a = 0$ at 
$K'_{\rm gap}$
and $K'_{\rm acc}$
and the detaching points at $K'_{\rm gap,h}$ and 
$K'_{\rm acc,l}$ as
%\yp{$K'$ here is $K'_{\rm acc,l}$?} as 
\begin{align} 
\frac{\dot{a}}{a} \simeq & \left(\frac{\dot{a}}{a} \right)_{\rm gap+dep,l} 
\left( \frac{\log K'_{\rm acc} - \log K'}{ \log K'_{\rm acc} - \log K'_{\rm acc,l} } \right) \nonumber \\
&  \ \ \ \ \ [ K'_{\rm acc,l} < K'  \lesssim K'_{\rm acc}] \\
\frac{\dot{a}}{a} \simeq & \left(\frac{\dot{a}}{a} \right)_{\rm gap,h} 
\left( \frac{\log K' - \log K'_{\rm gap}}{ \log K'_{\rm gap,h} - \log K'_{\rm gap} } \right) 
\nonumber \\ & \ \ \ \ \ [ K'_{\rm gap} \lesssim K'  < K'_{\rm gap,h}],
\end{align} 
where 
$(\dot{a}/{a})_{\rm gap+dep,l}$ and $(\dot{a}/{a})_{\rm gap,h}$ 
are $(\dot{a}/{a})_{\rm gap+dep}$ at $K' = K'_{\rm acc,l}$
and $(\dot{a}/{a})_{\rm gap}$ at $K' = K'_{\rm gap,h}$. 
As Fig.~\ref{fig:mig}a
clearly shows, simulation results converge to ``type II (gap)"
rather than ``type II (gap+dep)" in high $K'$ regimes,
because the gap is so deep there that the global disk depletion effect may be no longer applicable.
% (Section~\ref{sec:discussion}).
This formula (type II (full)) is plotted in Fig.~\ref{fig:mig_anly} by green lines.  
The formula consistently reproduces the \citet{Pan2025}'s results within a factor of 2 in broad ranges of parameters. 

\subsection{Concurrent growth and migration of gaseous planets}
\label{sec:grow_mig}

In order to demonstrate the impact of the outward migration on planet formation, we concurrently integrate the planetary growth by gas accretion,
\begin{align}
\frac{\dot{m}_{\rm p}}{m_{\rm p}} & = %\frac{1}{1+\zeta} 
\frac{\xi}{(1+\xi)}
\frac{\dot{m}_{\rm d}}{m_{\rm p}} 
\label{eq:growth}
\end{align}
and its migration with type II (full) formula in Section~\ref{sec:general}.
%In this case, it is convenient to rewrite Eq.~(\ref{eq:adot_a1}) as \begin{align} \frac{\dot{a}_{\rm I}}{a} = - \frac{25}{3} K' h \frac{\dot{m}_{\rm d}}{m_{\rm p}}. \end{align} \rev{Accordingly, Eqs.~(\ref{eq:adot_a2}) and (\ref{eq:outmigrate}) are rewritten.} Because this form has the same scaling factor $\dot{m}_{\rm d}/m_{\rm p}$ as Eq.~(\ref{eq:growth}), concurrent integration is easier \citep{Tanaka2020}. 
For planetary gas accretion, we also include the early phase of Kelvin-Helmholtz contraction limit into account \citep{Ida2018}.

\begin{figure*}[htbp]
\begin{center}
\includegraphics[width=160mm,angle=0]{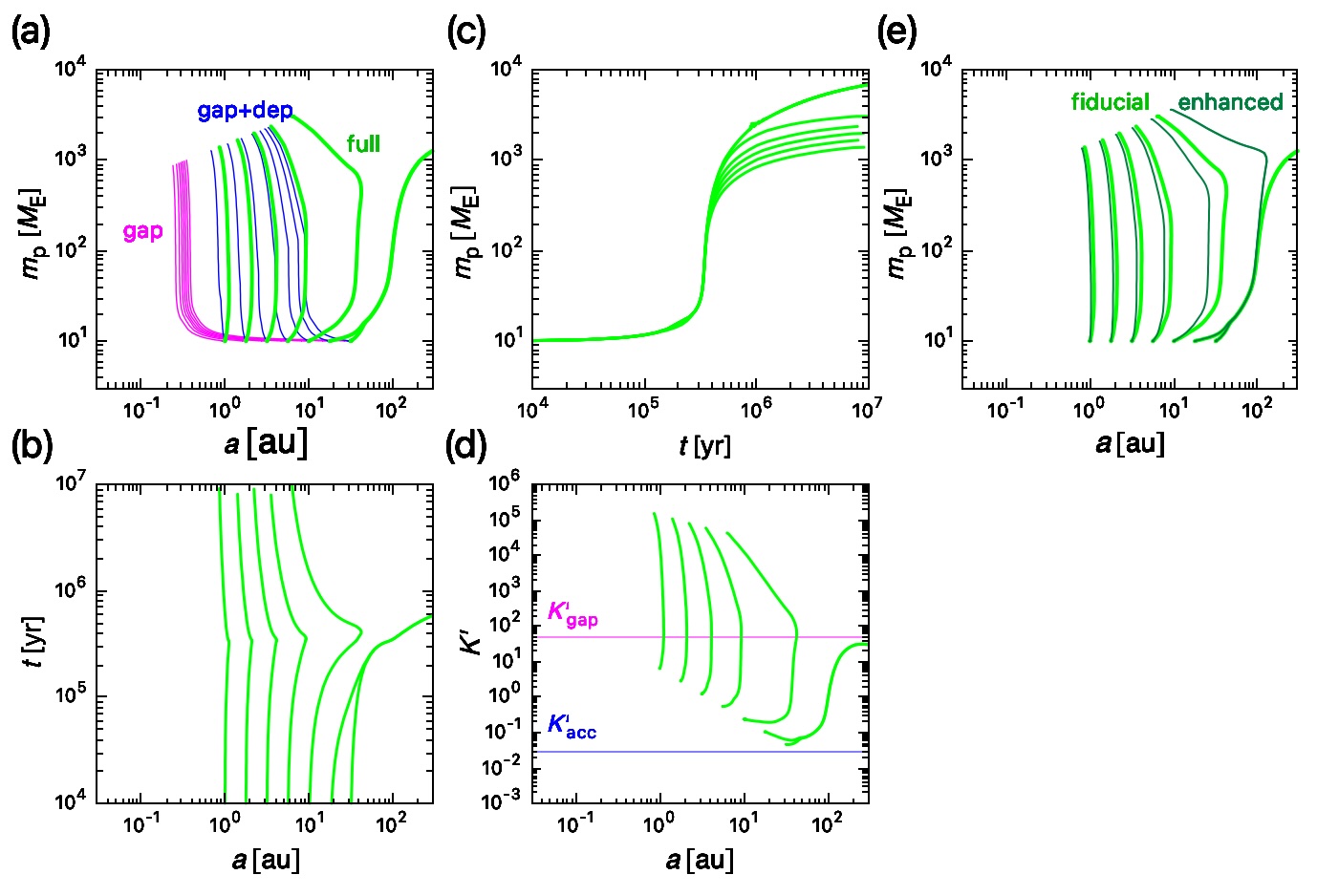}
\caption{\label{fig:evol_ma1} The mass and semimajor axis evolution of independent planets starting from $m_{\rm p} = 10 M_\oplus$ at 9 different locations of $a = 1$ - 30~au (Panel (a)).
For the disk, the self-similar solution with $\dot{m}_{\rm d} = 3 \times 10^{-8} (1 + t/t_{\rm diff})^{-3/2} M_\odot/{\rm yr} $ with $t_{\rm diff}=10^6\, \rm yrs$ and $T = 130\, (r/{1 \,\rm au})^{-1/2}\rm K$ is used.
The evolution with type II (full) formula in Section~\ref{sec:general} is shown in the green lines. For comparison, the evolutions with type II (gap) 
% formula (with only the gap effect) 
and type II (gap+dep) formulas
% (with the gap and the global disk depletion effects) 
are also shown in the magenta and blue lines.
The evolution is integrated until $t=10^7$ yrs.  
In Panels (b), (c), and (d), time evolution of the semimajor axis, the mass, and the gap parameter ($K'$) of individual planets are shown 
%in the case of 
with type II (full) formula.
In Panel (e), the result with enhanced Hill accretion rate by a factor 10 in type II (full) formula (dark green) is compared to the fiducial result (green). 
%\yp{is there a typo in the x-axis label of panel e?}
%In Panel \mg{(e)}, the fiducial result with the type II (full) formula (green lines) is compared to the result with enhanced gas accretion rate in Hill regime by a factor 10 (dark green lines).
}
\end{center}
\end{figure*}

Assuming the self-similar solution of a viscously expanding disk \citep{LydenBellPringle1974} where $T = 130\, (r/1 \, {\rm au})^{-1/2}\rm K$ and   $\dot{m}_{\rm d} = 3 \times 10^{-8} \, (1 + t/t_{\rm diff})^{-3/2} M_\odot/{\rm yr}$ with $t_{\rm diff}=10^6\, \rm yrs$, the mass and semimajor axis evolution of independent planets are shown in Fig.~\ref{fig:evol_ma1}.
We start integration from $10 M_{\oplus}$ planets in $3-30$ au. 
The interactions between planets are neglected in this particular calculations to highlight the outward migration, avoiding complex dynamics, although planetary interactions (resonance trapping, scattering, and collisions) were included in our previous population synthesis calculations \citep{IdaLin2010,Ida2013,Ida2018}. 

Figure~\ref{fig:evol_ma1}a shows comparison of results among type II (gap), (gap+dep), and (full) formulas. 
According to the hydrodynamical simulation results, we adopt type II (gap) formula at $K' > K'_{\rm gap}$ even in the calculation marked by type II (gap+dep). 

Comparison between type II (gap) and (gap+dep) formulas show that
the effect of the global disk depletion due to the planetary gas accretion significantly reduces inward migration of gas giants \citep[also see][]{Tanaka2020}.
It also shows that the outward migration due to planetary gas accretion brings planets further in particular in the outer region.
In the outermost region,
the outward migration
rather brings planets formed in $\gtrsim 20 \, \rm au$ to the outermost regions of the disk ($\gtrsim 300 \, \rm au$). 

Figures~\ref{fig:evol_ma1}b and c show runaway gas accretion and outward migration simultaneously occur around a few $\times 10^5$ yrs integration from $10\, M_\oplus$ cores.
During this phase, almost all the disk gas flow accretes onto the planet.

In Fig.~\ref{fig:evol_ma1}c, the evolution paths of $K'$ in individual planets are plotted. 
Because $K' \propto q^2 h^{-5}$ and $h \propto r^{1/4}$ in this calculation, $K'$ of each planet in inner regions is already high at $10\, M_\oplus$ and quickly increases as the planet growth to overshoot the outward migration regime of $K'_{\rm acc} < K' < K'_{\rm gap}$, and the migration turns into inward one.
%In this region, $K'$ becomes extremely high values and the classical type II migration that the planet migrates together with disk accretion may recover \citep{Chen2020}.

On the other hand, in outer regions, $K'$ is initially smaller and it is within the outward migration regime even if the planet acquires a Jupiter mass.
Therefore, the outward migration impacts planets in outer regions and it would significantly contribute to retain gas giants beyond 1 au.   
This also means that the runaway gas accretion lasts longer in outer regions (see panels a and b) and planets there can acquire larger final masses.

The planets starting at 18 and 30 au keep outward migration to reach the disk outermost regions of $300 \rm \, au$.
In this case, the increase in the planetary mass is compensated by the increase in $r$ and $h$, so that the increase in $K'$ is slower and the outward migration tends to be preserved.
If we set the radially exponential decay in the surface density in the self-similarity solution, the outward migration would stop at the decaying region.

Because this is converging migration, if planet-planet interactions are included, shepherding or strong scattering would happen.
It could potentially form HR8799-like systems, where multiple gas giants exist in almost circular orbits at $\sim 15$--70 au, while the giant planets around 1 au could also be formed in the same systems,
%\yp{implications for free floating planets?}
However, we note that the existence of planets that keep outward migration depends on disk conditions and actual gas accretion rates.

As pointed out in Section~\ref{sec:gas_acc}, hydrodynamical simulation results still have uncertainty in numerical factors of planetary gas accretion rate.
% in Hill accretion regime. 
\citet{Pan2025} suggests that the theoretical estimate 
in Eq.~(\ref{eq:Hill_acc2})
%adopted in this paper (Eq.~(\ref{eq:Hill_acc2})) 
may underestimate the accretion rate in Hill regime.
Figure~\ref{fig:evol_ma1}d shows the result with 10 times enhancement in the accretion rate in Hill regime in
% by a factor of 10 
type II (full) formula, as an extreme case, to be compared with the result with the fiducial type II (full) formula. 
Because the growth is accelerated, $K'$ tends to quickly escape from the outward migration regime and  
the planets that keep outward migration disappear.
They dissapear even with 3 times enhancement in the Hill accretion rate and also with that in the initial $\dot{m}_{\rm d}$.
%We also find that they dissapear if the initial $\dot{m}_{\rm d}$ is enhanced by a factor of 3.

Nevertheless, Figure~\ref{fig:evol_ma1}e shows that the mass-semimajor axis evolution is hardly affected by the factor 10 enhancement in 
%the planetary accretion rate in Hill regime
the Hill accretion rate, expect for the outermost planets.
This is because the planetary accretion rate has a cap by $\dot{m}_{\rm d}$. 
%Even if the accretion rate with infinite supply is enhanced, it cannot exceed the disk accretion rate. 
%It just slightly expands the planetary mass range of full accretion onto the planet and associated outward migration. 
Thus, the uncertainty in a factor of planetary gas accretion rate does not significantly affect the results in this paper. 

Finally, we point out that if some mechanism enhances accretion onto the planet in the deep gap (Section~\ref{sec:discussion}), the final planet distribution should be shifted further outward, compared with that predicted by the type II (gap+dep) formula.

\section{Discussion}
\label{sec:discussion}

%with high $K'\gtrsim K'_{\rm gap}$, 
%does not contribute to the total (disk+planet) mass flux 
In \citet{Li2024} and \citet{Pan2025}, $r_{\rm p}$ is fixed.
They estimated the migration from the torques from the $\Sigma$ distribution influenced by the planetary accretion.
Following their method, in this paper, we have approximated that $f_{\rm pass}+f_{\rm p}\simeq 1$.

As Fig.~\ref{fig:gas_acc} shows, when a planet grows and $q_{\rm th}$ exceeds $\sim 10$, a gap becomes so deep that $f_{\rm p}$ decreases as $q_{\rm th}$ increases.
With the assumption of $f_{\rm pass}+f_{\rm p}\simeq 1$, in this case,
$f_{\rm pass}$ must increase toward $\simeq 1$. 
However, the very deep gap should also prevent disk gas from crossing the gap, and
it may be more appropriate that $f_{\rm pass}+f_{\rm p} \ll 1$.

% We have pointed out a puzzle that $f_{\rm pass} \simeq 1$ in the very deep gap cases, within the current framework.

This inconsistency %with nearly unimpeded gas flow (implied by $f_{\rm pass} \simeq 1$) 
can only be resolved by either 
planetary migration or the breakdown of steady flow with mass accumulation 
(beyond $r_{\rm p}$) until the resumption of diffusion through the gap.
The mass accretion and the migration affect each other in a complicated manner \citep[for detailed discussion, see][]{Pan2025}.

\citet{Chen2020b} suggested that in the deep gap cases,
$f_{\rm pass} \ll 1$ and disk gas piles up outside the planetary orbit to push the planet inward, although they neglected the effect of planetary gas accretion.
%\yp{Did \citet{Chen2023} have such a conclusion? [Ida:Sorry, it is Chen et al. (2020)]}
On the other hand, \citet{Pan2025} suggest the eccentricity of the disk excited by the planet or other factors enhance accretion onto the planet in the deep gap cases. More detailed analysis on the deep gap cases is left to the future study.

% \begin{equation} {\dot m}_{\rm d} = {\dot m}_{\rm d,loc} + {\dot m}_{\rm p} - {\Gamma \over \Omega r_{\rm p}^2 } = (f_{\rm pass}+f_{\rm p}) \dot m_{\rm d} - m_{\rm p} {{\dot r}_{\rm p} \over 2 r_{\rm p}}, \end{equation}

% In addition to gas accretion, an embedded planet exerts a net (radially integrated) torque on the disk which induces its migration \citep{Lin1986, Lin1996}.  
% In a steady inward inflow from the outer boundary ($\dot{m}_{\rm d}$), the net inward mass flux (obtained from the continuity and angular momentum equations) is \begin{equation} {\dot m}_{\rm d} = {\dot m}_{\rm d,loc} + {\dot m}_{\rm p} - {\Gamma \over \Omega r_{\rm p}^2 } = (f_{\rm pass}+f_{\rm p}) \dot m_{\rm d} - m_{\rm p} {{\dot r}_{\rm p} \over 2 r_{\rm p}}, \end{equation} which is independent of $r$, where ${\dot m}_{\rm d,loc}$ is the local disk gas accretion rate. Gas is fed onto the outer regions of the disk at a rate ${\dot m}_{\rm d}$.
%In a steady state, ${\dot M}_{\rm net} = {\dot M}_{\rm infall}$. At $r \gg r_{\rm p}$ where the effect of the planet is neglected, ${\dot m}_{\rm d} \simeq {\dot m}_{\rm d,loc}$. %${\dot m}_{\rm d} \simeq {\dot m}_{\rm d,loc} + {\dot m}_{\rm p}$                 
%and evaluated $\dot{r}_{\rm p}$ afterwards from the resulted unperturbed global $\Sigma$ distribution with the gap opening effect.  
%at $r \lesssim r_{\rm p}$, 
%${\dot m}_{\rm d,loc}}, {\dot m}_{\rm p} \ll \dot{m}_{\rm d}$ (and \begin{equation}\dot{r}_{\rm p} \simeq - 2 {\dot m}_{\rm d} r_{\rm p} / m_{\rm p}.\end{equation}

\section{Conclusion}

To understand the observed mass and semimajor axis distribution of exoplanetary giant planets, type II orbital migration is a key process. 
%In particular, the observed pile-up of gas giant beyond 1 au has been one of the most important observed features that is difficult to be explained by conventional type II migration model.

Following \citet{Li2024}'s discovery of outward migration of a Jupiter-mass planet that accretes gas from the disk, 
%through high-resolution hydrodynamical simulations, 
\citet{Pan2025} performed parameter surveys to clearly show the broad parameter ranges in the disk and planetary mass 
for outward migration.

In this paper, 
we have discussed intrinsic physics of the outward migration and  
derived a general semi-analytical formula of the migration rate including outward migration.

The surface density of gas in the leading horseshoe-CPD region is higher than in the trailing horseshoe region, according to repeated mass loss from horseshoe flow to the planet during many-times horseshoe excursions.
%until disk gas inwardly passes the planetary orbit. 
This azimuthal asymmetry causes positive corotation torque, resulting in outward migration of the planet. If the planetary gas accretion effect is neglected, this azimuthal asymmetry never appears and the radial asymmetry drives the conventional inward migration. 
Therefore, whether the migration is inward or outward is regulated by competition between the azimuthal asymmetry by the corotation torque and the radial asymmetry mostly due to Lindblad torque (Eqs.~(\ref{eq:Lind}) and (\ref{eq:corot})).

We found that
%Because the gap width is broad enough to cover main Lindblad resonances and a circumplanetary disk (CPD) connected to horseshoe flow remains in the gap exerting corotation torque, 
the condition for the outward migration is regulated mostly by the gap depth: i) The gap is deep enough to remain the dominance of horseshoe corotation torque over Lindblad torque, and ii) it is shallow enough that the planet accretes most of disk gas accretion to maintain the azimuthal horseshoe asymmetry.
Based on the results of \citet{Pan2025}, 
we found that 
the outward migration condition is simpley given by
\begin{align}
& K'_{\rm acc} \lesssim K' \lesssim K'_{\rm gap} \: ; \\
& K'_{\rm acc} = 0.03 \,h_{0.05} \ ; \ K'_{\rm gap}=50,
\end{align}
where $K'$ is the gap depth parameter defined by $\Sigma_{\rm gap}/\Sigma_0 = 1/(1+K')$ and given by $K' = 1.28 \times 10^8 \, q^2 \, h_{0.05}^{-5} \, 
\alpha_{3}^{-1}$, 
$q = m_{\rm p}/M_*$, $h_{0.05} = h/0.05$, and $\alpha_3 = \alpha/10^{-3}$.
This condition covers broad parameter space.
For given $h$ and $\alpha$, it extends to the planet mass range of 
$1.5 \times 10^{-5} \, h_{0.05}^{3} \, \alpha_3^{1/2} \lesssim q \lesssim
 6.3 \times 10^{-4} \times h_{0.05}^{5/2} \, \alpha_3^{1/2}$.
 For smaller $\alpha$, for example, $\alpha=10^{-4}$, the range shifts to smaller values by a factor 3 from the above range. 
 Once a core exceeds the critical core mass ($\sim 3 \times 10^{-5}$) to start gas accretion, the migration is outward until it reaches $0.2 M_{\rm J}$. Because the migration of the low-mass planet is fast (the gap is still shallow) and the fast migration is outward, the impact on planet formation may be more significant for smaller $\alpha$.  

We also found that the absolute value of the outward migration rate is similar to that of the inward migration rate that neglects the azimutal asymmetry but takes account of global disk gas depletion due to the planetary gas accretion, although the migration direction is opposite. %\citep{Tanaka2020}.
Interpolating the outward and inward migration parameter regimes, we derived the general semi-analytical migration formula in Section~\ref{sec:general}.

Using the formula, we calculated evolution of concurrent growth and migration of planets starting with masses $10 \, M_\oplus$ to demonstrate that the outward migration plays an important role for the mass and semimajor axis distribution of gas giants, particularly in disk outer regions where the gap is shallower. 
It would potentially reproduce the observed pile-up of exoplanetary gas giants beyond 1 au, and could give an insight into formation of HR 8799-like systems. 
Detailed planet population synthesis calculations for quantitative comparison with the observed data of exoplanets will be done in a separate paper.

\acknowledgements

S.I. is supported by JSPS Kakenhi grant 21H04512 and 23H00143. Y.P.L is supported in part by the Natural Science Foundation of China (grants 12373070, and 12192223), the Natural Science Foundation of Shanghai (grant NO. 23ZR1473700). 

\bibliography{references.bib}{}
\bibliographystyle{aasjournalnolink}

\end{document}